\documentclass[twocolumn]{jpsj3}

\usepackage{color}

\def\gsim{\mathop {\vtop {\ialign {##\crcr 
$\hfil \displaystyle {>}\hfil $\crcr \noalign {\kern1pt \nointerlineskip } 
$\,\sim$ \crcr \noalign {\kern1pt}}}}\limits}
\def\lsim{\mathop {\vtop {\ialign {##\crcr 
$\hfil \displaystyle {<}\hfil $\crcr \noalign {\kern1pt \nointerlineskip } 
$\,\,\sim$ \crcr \noalign {\kern1pt}}}}\limits}

\title{
Theory for Intrinsic Magnetic Field in Chiral Superconductor Measured by $\mu$SR: 
Case of Sr$_2$RuO$_4$
}
\author{
Kazumasa {Miyake}$^{1,2}$ and Atsushi {Tsuruta}$^{3}$
}
\inst
{
$^{1}$
Center for Advanced High Magnetic Field Science, Graduate School of Science, \\
Osaka University, Toyonaka 560-0043, Japan\\
$^{2}$Toyota Physical and Chemical Research Institute, 41-1, Yokomichi, 
Nagakute, Aichi 480-1192, Japan\\
$^{3}$Division of Materials Physics, Department of Materials Engineering Science \\
Graduate School of Engineering Science, Osaka University, Toyonaka 560-8531, Japan
}

\recdate
{
\today
}

\abst
{
The local magnetic field induced by $\mu^{+}$ trapped at an interstitial site in  
a chiral superconductor with $p$-wave symmetry, such as Sr$_2$RuO$_4$, is discussed by solving the  
Bogoliubov-de Gennes equation on the two-dimensional square lattice. In the model Hamiltonian, 
the effect of the trapped $\mu^{+}$ extracting the electrons at surrounding Ru sites is 
phenomenologically taken into account as a non-magnetic impurity potential which locally destroys the 
chiral superconducting order with $p$-wave symmetry giving rise to local circulating current around 
$\mu^{+}$ site.  
It is shown that the size of the induced local magnetic field in the case with periodic boundary condition 
is far smaller compared to the case with open boundary condition without $\mu^{+}$, 
in which the surface current induced by 
destruction of superconducting order at the surface boundary gives contribution corresponding to 
the intrinsic angular momentum of the order of $\hbar N_{\rm s}/2$, with $N_{\rm s}$ being the number of 
superconducting electrons. 
This result qualitatively explains why the magnetic field $\simeq0.5$G measured by $\mu$SR in 
Sr$_2$RuO$_4$ is far smaller than the expected intrinsic magnetic field $\simeq 50$G  
which is nearly the same as the lower critical field $H_{{\rm c}1}\simeq 50$G.  
}

\begin{document}
\sloppy
\maketitle
\section{Introduction}

In the past decade, the problem concerning the intrinsic angular momentum (IAM) has revived as 
that of an intrinsic magnetic moment (IMM) in a spin-triplet chiral superconductor 
Sr$_2$RuO$_4$,~\cite{Maeno} in which the orbital part 
of superconducting gap is identified as 
\begin{equation}
\Delta_{\bf k}=\Delta\left[\sin(k_{x}a)+{\rm i}\sin(k_{y}a)\right],
\label{eq:1}
\end{equation}
where $a$ being the lattice constant in the two dimensional ab-plane~\cite{MiyakeNarikiyo}.  
This state is consistent with the temperature dependence of the specific heat 
(under the magnetic field),~\cite{Maeno}  
and theoretical investigations that suggest an importance of short range ferromagnetic correlations among 
quasiparticles~\cite{MiyakeNarikiyo,Hoshihara,Yoshioka}.  
This chiral state [Eq.\ (\ref{eq:1})] breaks the time-reversal symmetry (TRS), which is consistent 
with the report of a $\mu$SR measurement of a tiny but finite spontaneous magnetic field 
($\sim 0.5$G) around 
$\mu^{+}$ without external magnetic field~\cite{muSR}.  However, the size of this spontaneous 
magnetic field is far smaller than that expected from the IMM in the bulk system with the surface, 
as discussed below. 

If the IAM $L_{\rm in}$ is of the order of $N_{\rm s}\hbar/2$ and the gyro-magnetic ratio is given by 
$(-e/2m)$, with $e\,$($>0$) being the elementary charge, as in the classical case, 
the intrinsic magnetic moment (IMM) density $M_{\rm in}$ is estimated as 
\begin{equation}
M_{\rm in}\simeq -{n_{\rm s}\over 2}\mu_{0}{m\over m_{\rm band}^{\rm occ}}\mu_{\rm B},
\label{eq:10}
\end{equation} 
where $n_{\rm s}\equiv N_{\rm s}/V$, $\mu_{0}=4\pi\times 10^{-7}\,$H$\cdot$m$^{-1}$ is the magnetic 
permeability, $\mu_{\rm B}=e\hbar/2m$ is the Bohr magneton, and 
$m_{\rm band}^{\rm occ}$ is the harmonic average of band mass of electrons over occupied state in 
the Brillouin zone.~\cite{Tsuruta}  
Then, the magnetic flux density $B_{\rm in}$, 
without the external magnetic field H, is given by $M_{\rm in}$, because the relation 
$B=M+\mu_{0}H$ holds by its definition.~\cite{Purcell}  The electron number density $n$ of $\gamma$-band 
in Sr$_2$RuO$_4$, which is electron-like band,  is roughly estimated as 
\begin{equation}
n={1\over abc},
\label{eq:11}
\end{equation}
where $a=b=3.9\times 10^{-10}$m, and $c=(12.7/2)\times 10^{-10}$m is the length of edge of 
primitive cell of Sr$_2$RuO$_4$ along $a$ ($b$) and $c$ directions, respectively~\cite{Mackenzie}. 
The magnetization density $M_{\rm in}$ is given by the relation
\begin{equation}
M_{\rm in}=-\mu_{0}{e\over 2m_{\rm band}^{\rm occ}}L_{\rm in}=
-{\hbar \over 2}n\mu_{0}{e\over 2m_{\rm band}^{\rm occ}},
\label{eq:12}
\end{equation} 
where $m_{\rm band}\simeq 2.9\,m$ is the effective mass of $\gamma$-band of 
Sr$_2$RuO$_4$~\cite{Mackenzie}.  
Therefore, the intrinsic magnetic flux density $B_{\rm in}$ is estimated 
as
\begin{eqnarray}
& &B_{\rm in}=-{10^{-30}\over abc}\,{m\over m_{\rm band}^{\rm occ}}\times 5.8\, {\rm T}
\nonumber
\\
& &\qquad
\simeq -2.1\times 10^{-2}\, {\rm T}=-2.1\times 10^{2}\, {\rm G}.
\label{eq:13}
\end{eqnarray}
This value is larger than the ``observed" lower critical field 
$B^{\rm obs}_{{\rm c}1}=5.0\times10^{-3}\,{\rm T}$ of 
Sr$_2$RuO$_4$~\cite{Akima}.  
However, since Sr$_2$RuO$_4$ has other two bands, hole-like $\alpha$-band and 
electron-like $\beta$-band, a considerable cancellation in the IMM is expected among 
electron-like $\beta$- and $\gamma$-band and hole-like $\alpha$-band. 
{Indeed, 
the size of $B_{\rm in}$ decreases to $|B_{\rm in}|\lsim 5.0\times 10^{-3}\,{\rm T}$~\cite{comment}  
which is comparable to the ``observed'' lower critical field 
$B^{\rm obs}_{{\rm c}1}=5.0\times10^{-3}\,{\rm T}$. Therefore, the actual $B_{\rm in}$ in 
Sr$_2$RuO$_4$ is expected to be almost screened out by the Meissner effect. } 

Then, it is reasonable to consider that the spontaneous magnetic field ($\sim 0.5$G) measured by 
$\mu$SR~\cite{muSR} is not related to the bulk IMM but to other physical mechanism.  
One of possible ideas for this is that the positive charge of $\mu^{+}$ attracts electrons on 
the Ru site adjacent to stopping $\mu^{+}$, which acts as a non-magnetic impurity potential 
destroying the superconductivity gap given by Eq.\ (\ref{eq:1}) there,~\cite{Onishi,SchmittRink,Hirschfeld} 
resulting in the local electric current surrounding $\mu^{+}$. 
Namely, the cancellation of relative rotation of Cooper pairs becomes incomplete there, giving rise 
to a circulating current around the position of the impurity, i.e., the stopping site of $\mu^{+}$, 
and local magnetic flux density (magnetic field) $B_{\rm loc}$ which causes the $\mu$ spin 
rotation ($\mu$SR). However, it is a nontrivial problem whether this induced $B_{\rm loc}$ 
can be smaller than the $B_{\rm in}$ 
induced by the surface current of the system if the impurity potential is strong enough to suppress 
the superconducting gap adjacent to the impurity, while the  $B_{\rm loc}$ is expected to be smaller than 
the  $B_{\rm in}$ if the impurity potential is moderate comparable to the pairing interaction. 

The purpose of the present paper is to clarify this problem by solving the 
Bogoliubov-de Gennes equation on the two-dimensional square lattice model 
with the inter-site attractive interaction causing the chiral superconductivity given by Eq.\ (\ref{eq:1}) and 
the effect of $\mu^{+}$ on the electrons at surrounding sites.  
Organization of this paper is as follows.  In Sect.\ 2, we introduce the model on the square lattice 
with attractive interaction between nearest neighbor sites and the effect of the repulsive impurity potential 
at the sites adjacent to stopping $\mu^{+}$. 
In Sect.\ 3, we discuss a formalism for explicit calculations. In Sect.\ 4, we present  
the results of magnetic flux density at $\mu^{+}$ site and 
the pattern of electric current induced around the $\mu^{+}$ site. 
Finally, in Sect.\ 5, the relation between the numerical results and the spontaneous magnetic field 
$\simeq 0.5\,$G observed by $\mu$SR in Sr$_2$RuO$_4$ is discussed, and perspective of the present 
results is discussed in relation to the fact that spontaneous magnetic field is observed in a series of 
superconductors with crystal structures without inversion center.   

\section{Effect of $\mu^{+}$ in Chiral Superconductor on Square Lattice}
\subsection{Model Hamiltonian}
\begin{figure}[h]
\begin{center}
\rotatebox{0}{\includegraphics[width=0.7\linewidth]{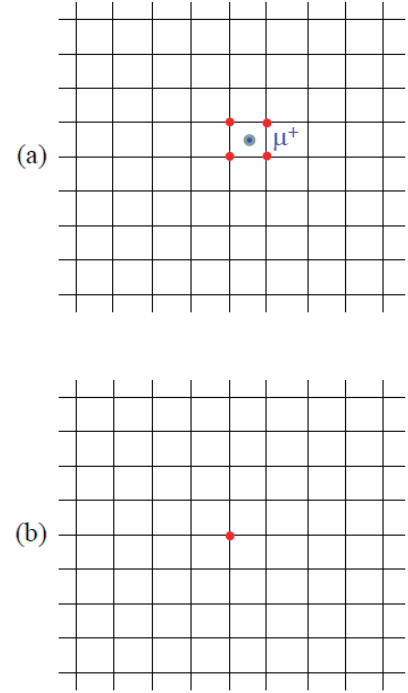}}
\caption{(a) $\mu^{+}$ (blue large circle) and extra electrons attracted on Ru sites adjacent to 
$mu^{+}$ (red small circles). 
(b) Simplified model in which extra electrons are attracted on one Ru site. 
}
\label{MuonSite}
\end{center}
\end{figure}

In order to study the effect of a $\mu^{+}$ stopping in the chiral superconductor on two-dimensional lattice, 
a model of Sr$_2$RuO$_4$, we start with the following Hamiltonian 
\begin{eqnarray}
& &
{\cal H}=-\mu\sum_{i\sigma}c^{\dagger}_{i\sigma}c_{i\sigma}
-t\sum_{\langle i,j\rangle\sigma}c^{\dagger}_{i\sigma}c_{j\sigma}
\nonumber
\\
& &
\qquad
-{V\over 2}\sum_{\langle i,j\rangle\sigma}
c^{\dagger}_{j\sigma}c^{\dagger}_{i{\bar \sigma}}c_{i{\bar \sigma}}c_{j\sigma}
+U\sum_{\sigma}c^{\dagger}_{{\rm O}\sigma}c_{{\rm O}\sigma},
\label{eq:CS1}
\end{eqnarray}
where $\mu$, $t$, and $V$ are the chemical potential, the transfer integral between nearest neighbor 
(n.n.) sites of the square lattice, and the attractive interaction between electrons at n.n. sites, 
respectively, and $c^{\dagger}_{i\sigma}$ ($c_{i\sigma}$) is the creation 
(annihilation) operator of electron at $i$-th site with spin component $\sigma$ ($=\uparrow$ 
or $\downarrow$).  The symbol $\langle i,j\rangle$ indicates the summation is taken over the n.n. sites.  
The last term in Eq.\ (\ref{eq:CS1}) represents the repulsive impurity potential $U$ at 
the origin of the lattice ($i={\rm O}$) which simulates 
the effect of electrons attracted on Ru site near the $\mu^{+}$ stopping at interstitial position in 
Sr$_2$RuO$_4$, as shown in Fig.\ \ref{MuonSite}(a). Here, we have simplified the effect of $\mu^{+}$ 
as Eq.\ (\ref{eq:CS1}) in which the position of $mu^{+}$ is shifted on the Ru site, as shown in 
 Fig.\ \ref{MuonSite}(b), for the sake of simplicity of numerical calculations. 

Hereafter, we consider the spin triplet paring with $S_{z}=0$, and introduce a superconducting gap 
$\Delta_{ij}$ in the spin-triplet manifold as 
\begin{equation}
\Delta_{ij}={V\over 2}\langle c_{i\uparrow}c_{j\downarrow}+
c_{i\downarrow}c_{j\uparrow}\rangle, 
\label{eq:CS2}
\end{equation} 
where $\langle\cdots\rangle$ means the average by the mean-field Hamiltonian 
$H_{\rm mf}$ given as 
\begin{eqnarray}
& &
{\cal H}_{\rm mf}=-\mu\sum_{i\sigma}c^{\dagger}_{i\sigma}c_{i\sigma}
-t\sum_{\langle i,j\rangle\sigma}c^{\dagger}_{i\sigma}c_{j\sigma}
\nonumber
\\
& &
\qquad
+\sum_{\langle i,j\rangle}\left\{\left[\Delta_{ij}
(c^{\dagger}_{j\uparrow}c^{\dagger}_{i\downarrow}+c^{\dagger}_{j\downarrow}c^{\dagger}_{i\uparrow})
+{\rm h.c.}\right]-\frac{|\Delta_{ij}|^{2}}{V}\right\}
\nonumber
\\
& &
\qquad
+U\sum_{\sigma}c^{\dagger}_{{\rm O}\sigma}c_{{\rm O}\sigma}.
\label{eq:CS3}
\end{eqnarray}
Here the gap $\Delta_{ij}$ depends on lattice sites $i$ and $j$ in general, and its dependence 
is determined self-consistently by solving the Bogoliubov-de Gennes equation (of lattice version) 
together with the relation (\ref{eq:CS2})~\cite{Onishi}.  
The gap $\Delta_{ij}$ is odd with respect to the interchange of $i\rightleftharpoons j$: 
\begin{equation}
\Delta_{ij}=-\Delta_{ji},
\label{eq:CS2a}
\end{equation}
which manifests the odd-parity pairing.  
Note that, in the case of uniform system without boundary, the stablest gap of those given 
by Eq.\ (\ref{eq:CS2}) is expressed in a wave-vector 
representation as Eq.\ (\ref{eq:1}). 

\subsection{Magnetic field $B_{z}$ at $\mu^{+}$ site in band picture}
Similar approximation is adopted for the integral along the $y$-direction.  
As shown in Ref.\ \citen{Tsuruta}, 
the magnetization operator ${\hat M}_{z}$ due to orbital motion is given by 
\begin{equation}
{\hat M}_{z}=\mu_{0}{(-e)\over 2m_{\rm b}}\sum_{i}({\bf r}_{i}\times{\bf p}_{i})_{z},
\label{eq:CS11}
\end{equation}
where the ``momentum" operator ${\bf p}_{i}$ at the $i$-th site is defined by 
\begin{eqnarray}
& &
p_{xi}\equiv {-{\rm i}\over 2}{\hbar\over a}\sum_{\sigma}
\left[ (c^{\dagger}_{(i_{x}+1,i_{y}){\sigma}}-c^{\dagger}_{(i_{x}-1,i_{y}){\sigma}})c_{(i_{x},i_{y}){\sigma}}
\right.
\nonumber
\\
& &
\left.
\qquad\qquad
-c^{\dagger}_{(i_{x},i_{y}){\sigma}}(c_{(i_{x}+1,i_{y}){\sigma}}-c_{(i_{x}-1,i_{y}){\sigma}})\right]
\nonumber
\\
& &
p_{yi}\equiv {-{\rm i}\over 2}{\hbar\over a}\sum_{\sigma}
\left[ (c^{\dagger}_{(i_{x},i_{y}+1){\sigma}}-c^{\dagger}_{(i_{x},i_{y}-1){\sigma}})c_{(i_{x},i_{y}){\sigma}}
\right.
\nonumber
\\
& &
\left.
\qquad\qquad
-c^{\dagger}_{(i_{x},i_{y}){\sigma}}(c_{(i_{x},i_{y}+1){\sigma}}-c_{(i_{x},i_{y}-1){\sigma}})\right].
\label{eq:CS9}
\end{eqnarray}
The relation (\ref{eq:CS11}) is a band-version of conventional form with gyro-magnetic ratio 
$(-e/2m_{\rm b})$, where  $m_{\rm b}\equiv \hbar^{2}/2ta^{2}$ is the band mass at 
$\Gamma$-point. The above definition of $m_{\rm b}$ corresponds to the free electron like 
dispersion of tight binding dispersion around the $\Gamma$-point, $(k_{x},k_{y})=(0,0)$.  
Namely, 
\begin{eqnarray}
& &
\epsilon_{k}=-2t(\cos\,k_{x}a+\cos\,k_{y}a)
\nonumber
\\
& &
\quad\,
\simeq -4t+ta^{2}(k_{x}^{2}+k_{y}^{2})+\cdots\,.
\label{eq:CS13}
\end{eqnarray} 

Corresponding to the relation (\ref{eq:CS11}), ${\hat B}_{z}(0,0)$, 
the operator for the $z$-component of the local magnetic flux density vector at the center of the crystal 
lattice, ${\bf r}_{\rm O}\equiv(0,0)$, is given by a lattice version of  the Biot-Savart law~\cite{Purcell} 
as follows:
\begin{equation}
{\hat B}_{z}(0,0)=\frac{\mu_{0}}{4\pi}{(-e)\over m_{\rm b}}\sum_{i}
\frac{({\bf r}_{i}\times{\bf p}_{i})_{z}}{|{\bf r}_{i}|^{3}}. 
\label{eq:CS14}
\end{equation}

\section{Formalism of Numerical Calculations}
An explicit form of the Bogoliubov-de Gennes equation for the mean-field Hamiltonian (\ref{eq:CS3}) 
with the superconducting gap of $S_{z}=0$, Eq.\ (\ref{eq:CS2}), is given by~\cite{deGennes2}
\begin{eqnarray}
& &\varepsilon\, u_{i}=-\mu \,u_{i}-t\,u_{j}+\sum_{\langle j,i\rangle}\Delta_{ij}v_{j}+Uu_{i}\delta_{i{\rm O}},
\label{eq:R1}
\\
& &\varepsilon\, v_{i}=\mu \,v_{i}+t\,v_{j}+\sum_{\langle j,i\rangle}\Delta_{ij}^{*}u_{j}+Uu_{i}\delta_{i{\rm O}},
\label{eq:R2}
\end{eqnarray}
where $\delta_{ij}$ is the Kronecker delta. By solving these equations and the superconducting gap 
[Eq.\ (\ref{eq:CS2})] self-consistently, the average of the spontaneous magnetic field at $\mu^{+}$ site 
[Eq.\ (\ref{eq:CS14})] is obtained. 

An actual calculation is 
performed as follows. Hereafter, we focus our discussion in the half-filled case. Equations 
(\ref{eq:R1}) and (\ref{eq:R2}) are diagonalized by means of a unitary transformation ${\cal U}$ 
to give the mean-field Hamiltonian 
\begin{equation}
H_{\rm mf}=\sum_{m=1}^{N_{\rm L}}\varepsilon_{m}\gamma^{\dagger}_{m\uparrow}\gamma_{m\uparrow}
+\sum_{m=1}^{N_{\rm L}}(-\varepsilon_{m})\gamma^{\dagger}_{m\downarrow}\gamma_{m\downarrow},
\label{eq:R3}
\end{equation}
where $N_{\rm L}$ is the number of lattice sites, $0\le\varepsilon_{1}\le\varepsilon_{1}\dots
\le\varepsilon_{N_{\rm L}}$, and the fermion operators $\gamma$ describing quasiparticles 
are related to the electron operators $a$ by 
\begin{eqnarray}
& &[c^{\dagger}_{1\uparrow}, \cdots,\,c^{\dagger}_{N_{\rm L}\uparrow},\,
c_{1\downarrow}, \cdots,\,c_{N_{\rm L}\downarrow}]
\nonumber 
\\
& &\qquad
=[\gamma^{\dagger}_{1\uparrow},\cdots,\, \gamma^{\dagger}_{N_{\rm L}\uparrow},\, 
\gamma_{1\downarrow},\cdots,\, \gamma_{N_{\rm L}\downarrow}]{\cal U}^{\dagger}. 
\label{eq:R4}
\end{eqnarray}
Substituting Eq.\ (\ref{eq:R4}) into Eq.\ (\ref{eq:CS2}), we obtain the self-consistent 
equation for the gap $\Delta_{ij}$ as 
\begin{eqnarray}
& &
\Delta_{ij}=\frac{V}{2}\sum_{m=1}^{N_{\rm L}}\left[
({\cal U})^{*}_{j+N_{\rm L},m}({\cal U})_{i,m}
-({\cal U})^{*}_{i+N_{\rm L},m}({\cal U})_{j,m}\right]
\nonumber
\\
& &
\qquad\qquad\qquad\qquad\qquad\qquad\qquad
\times[1-f(\varepsilon_{m})]
\nonumber
\\
& &
\qquad
+\frac{V}{2}\sum_{m=1}^{N_{\rm L}}\left[
({\cal U})^{*}_{j+N_{\rm L},m+N_{\rm L}}({\cal U})_{i,m+N_{\rm L}}
\nonumber
\right.
\\
& &
\left.
\qquad\qquad\quad\,\,
-({\cal U})^{*}_{i+N_{\rm L},m+N_{\rm L}}({\cal U})_{j,m+N_{\rm L}}\right]
f(\varepsilon_{m}), 
\label{eq:R5}
\end{eqnarray}
where ${\cal U}$ depends on $\Delta_{ij}$'s and $\varepsilon_{m}$ ($m=1,\cdots, N_{\rm L}$), 
and $f(x)$ is the Fermi distribution function $f(x)\equiv(e^{x/T}+1)$.    

\begin{figure}[h]
\begin{center}
\rotatebox{0}{\includegraphics[width=0.5\linewidth]{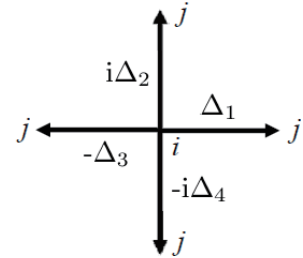}}
\caption{Phase of superconducting gap $\Delta_{ij}$ in clean system under periodic boundary 
condition with $i$-site chosen as a center. 
}
\label{Gap_Phase}
\end{center}
\end{figure}

We have solved Eqs.\ (\ref{eq:CS2}), (\ref{eq:CS3}), and (\ref{eq:R3}) $\sim$ (\ref{eq:R5}) 
self-consistently using the numerical diagonalization method and obtained the gap 
$\Delta_{ij}$'s and the energy level $\varepsilon_{m}$ ($m=1,\cdots, N_{\rm L}$).  
Numerical calculations have been performed for the square lattice of sizes $N_{\rm L}=20\times 20$ and 
$N_{\rm L}=30\times 30$ with the periodic boundary condition because we are considering the case 
without the effect of boundary surface of the system.  
In the pure system with periodic boundary condition, the phase of superconducting gap $\Delta_{ij}$ 
can be chosen as shown in Fig.\ \ref{Gap_Phase} and $\Delta_{i}$ ($i=1\sim4$) are independent of the 
site index $i$. However, in the system with an impurity, the gap functions $\Delta_{ij}$ do not have such 
simple form and should be determined self-consistently.  

\section{Magnetic Flux Density and Current Pattern}
\begin{figure}[h]
\begin{center}
\rotatebox{0}{\includegraphics[width=0.8\linewidth]{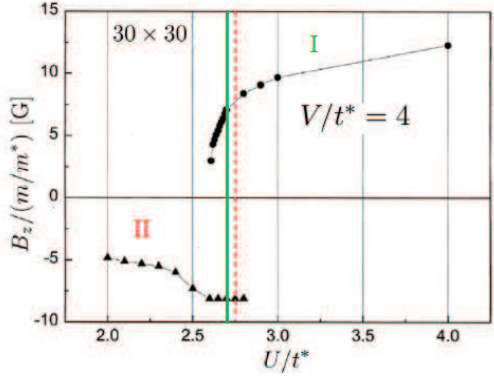}}
\caption{
Relation between the spontaneous magnetic flux density $B_{z}$ at the origin ($\mu^{+}$ site) and 
the impurity potential  $U/t^{*}$ at the origin. $m$ and $m^{*}$ are 
mass of free electron and the effective mass renormalized by the correlation effect, respectively. 
Lattice size is $N_{\rm L}=30\times 30$. The pair interaction is set as $V/t^{*}=4$，where 
$t^{*}$ is the effective hoping of quasiparticles renormalized by the correlation effect. 
At $U/t^{*}\simeq 2.75$ (shown by vertical dashed line), the solutions changes from type I to type II. 
Current patterns at $U/t^{*}=2.7$ (shown by vertical solid line), are shown in Fig.\ \ref{CurrentPattern}. 
}
\label{IMM_vs_U}
\end{center}
\end{figure}

Figure\ \ref{IMM_vs_U} shows the dependence of the spontaneous magnetic flux density $B_{z}$ 
at the origin ($\mu^{+}$ site) on the impurity potential  $U/t^{*}$ for the case that the pair interaction 
is given by $V=4t^{*}$, where $t^{*}$ is the effective hopping of quasiparticles renormalized by 
correlation effect and $m/m^{*}$ is the ratio of free electron mass and the effective mass renormalized 
by correlation effect.  The lattice size is taken as $N_{\rm L}=30\times 30$. 
There exist two solutions, I and II, 
which make the accuracy of self-consistency stationary as ${\cal O}(10^{-3})$ corresponding to the value 
of $U/t^{*}$. At $U/t^{*}\gsim 2.75$, the solution with highest accuracy is the type I, 
while that at  $U/t^{*}\lsim 2.75$e is the type II. These two solutions exhibit first order like transition at 
$U/t^{*}\simeq 2.75$ shown by vertical dashed line, 
and there exist metastable solutions around $U/t^{*}\simeq 2.75$. 

\begin{figure}[h]
\begin{center}
\rotatebox{0}{\includegraphics[width=1.0\linewidth]{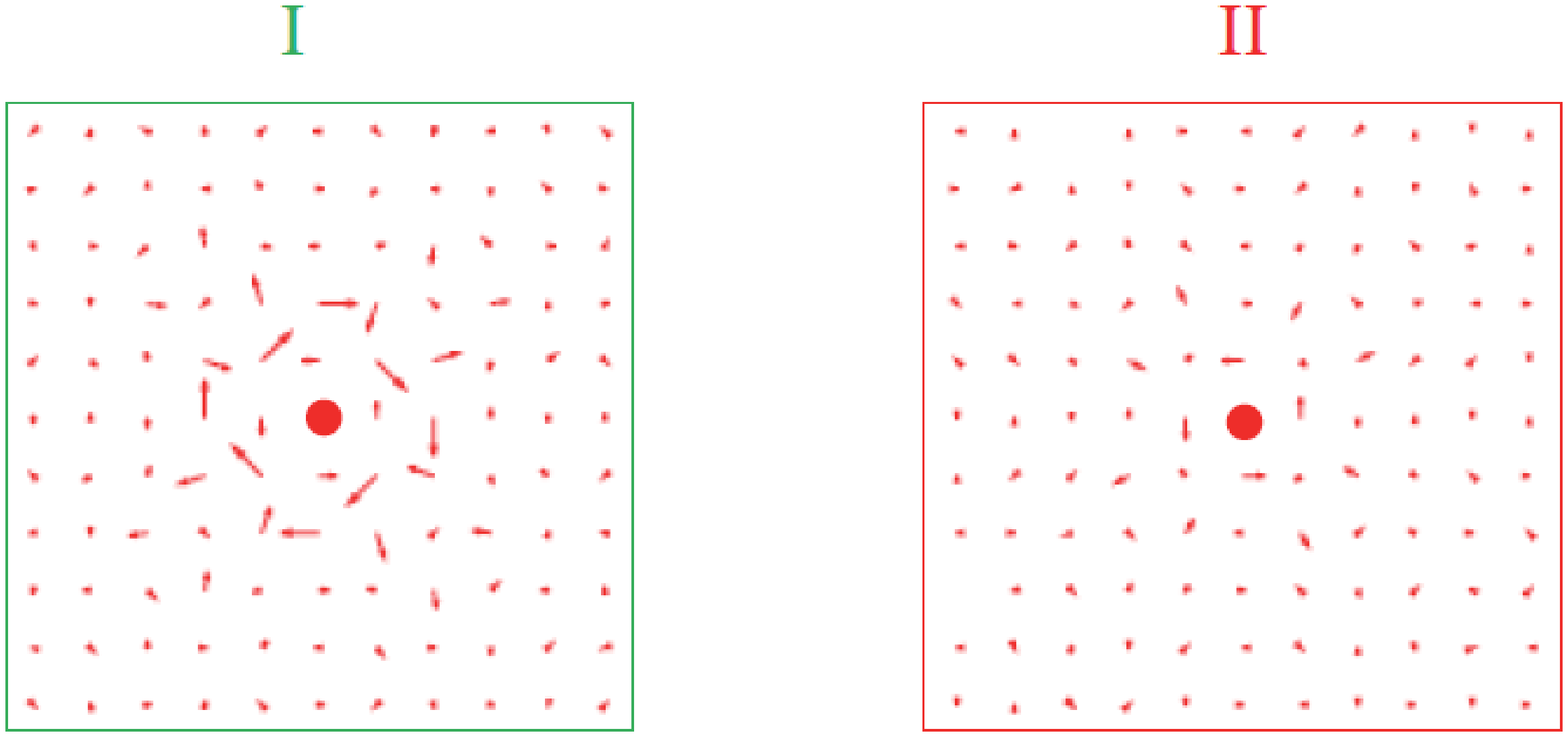}}
\caption{
Patterns of current around the impurity site ($\mu^{+}$ site shown by filled circle) 
for tow types of solutions I and II 
for $U/t^{*}=2.7$ (corresponding to the vertical solid line shown 
in Fig.\ \ref{IMM_vs_U}). 
}
\label{CurrentPattern}
\end{center}
\end{figure}

Figure\  \ref{CurrentPattern} shows the current pattern for $U/t^{*}=2.7$ 
(shown by vertical solid line in Fig.\ \ref{IMM_vs_U}) for the type I and type II.
Spontaneous magnetic field of the type I is $B_{z}>0$, and that for II is $B_{z}<0$. 
This is understood from the direction of the current. Namely, it is clockwise around the 
impurity ($\mu^{+}$) for the type I so that the magnetic field points to the positive direction 
of $z$-axis, while it is counter clockwise for the type II so that the direction of the magnetic field 
is opposite.  The important point is that, in both cases, the magnitudes of the magnetic field induced at 
$\mu^{+}$ site are given by   
\begin{equation}
|B_{z}(0,0)|\sim 10\times\frac{m}{m^{*}}\,{\rm G}. 
\label{eq:B_{z}}
\end{equation}
Since $m^{*}/m\sim 10$ in Sr$_2$RuO$_4$\cite{Mackenzie}，the induced magnetic field is expected 
to be the order of $1\,{\rm G}$.  This value of $B_{z}(0,0)$ is the same order as the spontaneous 
magnetic field observed by $\mu$SR~\cite{muSR}, explaining the extremely small magnetic field 
observed by the $\mu$SR measurement.  

Note that this spontaneous magnetic field at $\mu^{+}$ site is not screened by the Meissner effect 
because it is the magnetic field in the region apart from the $\mu^{+}$ site by the penetration 
depth $\lambda$ ($\sim13{\rm nm}$ in Sr$_2$RuO$_4$\cite{Mackenzie}） that is screened by the 
Meissner effect.

\begin{figure}[h]
\begin{center}
\rotatebox{0}{\includegraphics[width=0.8\linewidth]{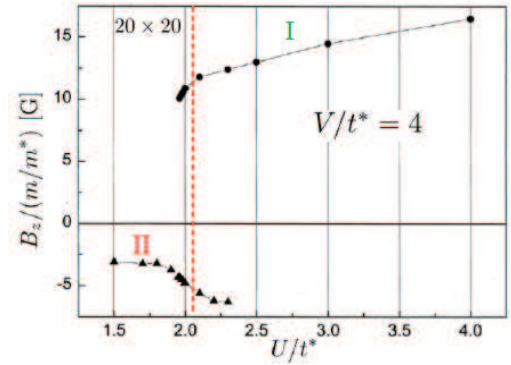}}
\caption{
Relation between the spontaneous magnetic flux density $B_{z}$ at the origin ($\mu^{+}$ site) and 
the impurity potential  $U/t^{*}$ at the origin for the lattice size $N_{\rm L}=20\times 20$. 
At $U/t^{*}\simeq 2.10$ (shown by the vertical dashed line) the stable solution changes from type I to 
type II.}
\label{IMM_vs_U_20x20}
\end{center}
\end{figure}

Figure \ref{IMM_vs_U_20x20} shows the results corresponding to Fig.\ \ref{IMM_vs_U} for the lattice size 
$N_{\rm L}=20\times 20$. A general tendency is fundamentally the same as that shown in 
Fig.\ \ref{IMM_vs_U} for$N_{\rm L}=30\times 30$.  However, the critical value of $U_{\rm cr}/t^{*}$ giving the 
transition between two types I and II shifts from $U_{\rm cr}/t^{*}\simeq 2.75$ to the lower value 
$U_{\rm cr}/t^{*}\simeq 2.10$.  This may be interpreted as an interference effect of 
two impurities the effect of which inevitably appears due to adopting the periodic boundary condition. 
In this sense, the calculations with much larger lattice size are desired, which are left for future study. 

{
Concluding this section, let us briefly discuss how the results on the size of the spontaneous 
magnetization depends on the strength of the intersite attractive interaction $V$. 
According to Ref.\ \citen{Tsuruta}, the extent $\xi^{*}$ of the Cooper pair in the low temperature 
limit ($T\ll T_{\rm c}$) is estimated as $\xi^{*}/a\simeq 2.6$. On the other hand, $\xi^{*}=\pi\xi_{0}$ of 
Sr$_2$RuO$_4$ is estimated as $\xi^{*}/a\simeq 5.3\times 10^{2}$.~\cite{Mackenzie}  
As shown in Appendix, a factor 
$\sum_{i}({\bf r}_{i}\times{\bf p}_{i})_{z}/|{\bf r}_{i}|^{3}$ in Eq.\ (\ref{eq:CS14}) is estimated as 
\begin{equation}
\sum_{i}\frac{|({\bf r}_{i}\times{\bf p}_{i})_{z}|}{|{\bf r}_{i}|^{3}}
\approx \frac{2\pi p_{0}}{a^{2}}\left(\ln\frac{\xi^{*}}{a}+\gamma\right)
e^{-a/\xi^{*}},
\label{Biot-Savart}
\end{equation}
where $p_{0}$ is the size of momentum at the nearest neighbor site around the origin (impurity site) and 
$\gamma\simeq 0.557\cdots$ is the Euler constant. Namely, this factor has only weak logarithmic 
dependence of $\xi^{*}/a$ in the region $\xi^{*}\gg a$, so that a huge ratio of $\xi^{*}$ between those of 
the present model and Sr$_2$RuO$_4$, $5.3\times 10^{2}/2.6\simeq 2.0\times 10^{2}$, gives 
a difference only of a factor 5. 
}
\section{Summary and Perspective}
We have clarified the origin of extremely small spontaneous magnetic field of $B\simeq 0.5\,$G 
observed in a p-wave chiral superconductor Sr$_2$RuO$_4$ by $\mu$SR measurement~\cite{muSR} 
on the basis of 
numerical analysis of the model Hamiltonian on the square lattice with the nearest-neighbor 
attraction with the effect of excess electrons on the lattice point which are 
attracted by the $\mu^{+}$ itself stopped in interstitial of the lattice. 
The crucial point was that the excess electrons attracted around the $\mu^{+}$ work to destroy the 
chiral superconducting order around them and in turn manifests the circulating currents around the 
$\mu^{+}$. This is in marked contrast with the case without $\mu^{+}$ in which the currents associated 
with chiral motion of the Cooper pairs are canceling  with each other in the bulk system 
except near the system boundary.~\cite{Tsuruta}   

The time-reversal-symmetry breaking mechanism discussed in the present paper is also different 
form that cause by the effect of spin space in the equal spin paring state of spin triplet 
paring~\cite{Miyake} which was
discussed in relation to the excess Knight shift increase below the superconducting transition 
temperature observe in Sr$_2$RuO$_4$.~\cite{Ishida} 

The model and theory developed in the present paper is possibly related to origins of 
phenomena of spontaneous time-reversal-symmetry breaking with small intrinsic magnetic fields 
{of the order of $1\,$G which are systematically} observed by the $\mu$SR measurement 
in a series of exotic superconductors,  
 (U;Th)Be$_{13}$,~\cite{(U;Th)Be13} 
UPt$_3$,\cite{UPt3} 
(Pr;La)(Os;Ru)$_4$ Sb$_{12}$,~\cite{(Pr;La)(Os;Ru)4Sb12} 
LaNiC$_2$,~\cite{Hillier2}  
PrPt$_4$Ge$_{12}$,~\cite{PrPt4Ge12} 
LaNiGa$_2$,~\cite{Hillier} 
Re$_6$Zr,~\cite{Re6Zr} and
Lu$_5$Rh$_6$Sn$_{18}$,~\cite{Bhattacharyya} {and so on.}

\section*{Acknowledgments}
This work is supported by Grants-in-Aid for Scientific 
Research (No. 17K05555) from the Japan Society for the Promotion of Science. 
One of us (K.M.) is grateful to Jorge Quintanilla for directing our attention to the 
spontaneous magnetic field observed by $\mu$SR experiments, especially in 
a series of superconductors with and without inversion center of the crystal, which was crucial 
for us to think seriously the case of Sr$_2$RuO$_4$, and for the hospitality at the University of Kent 
where the final stage of this work has been performed through the EPSRC project ”Unconventional 
superconductors: New paradigms for new materials” 
(grant references EP/P00749X/1 and EP/P007392/1).

{
\appendix
\section{Cooper-Pair Size Dependence of Biot-Savart Contribution}
In this appendix, we estimate the size of 
$\sum_{i}({\bf r}_{i}\times{\bf p}_{i})_{z}/|{\bf r}_{i}|^{3}$ in Eq.\ (\ref{eq:CS14}). In the case of 2-dimensional 
lattice with the lattice constant $a$, the summation with respect to sites is approximated by 
integration in the 2-dimensional space as follows:

\begin{eqnarray}
& &
\sum_{i}
\frac{{\bf r}_{i}\times{\bf p}_{i}}{|{\bf r}_{i}|^{3}}
\simeq \frac{1}{a^{2}}\int{\rm d}{\bf r}\frac{{\bf r}\times{\bf p}({\bf r})}{r^{3}}
\nonumber
\\
& &
\sim \frac{2\pi}{a^{2}}\int_{b}^{\infty}{\rm d}r\,\frac{p_{0}\,e^{-r/\xi^{*}}}{r}
\nonumber
\\
& &
=\frac{2\pi p_{0}}{a^{2}}\int_{a/\xi^{*}}^{\infty}{\rm d}x \frac{e^{-x}}{x}
\nonumber
\\
& &
\approx
\frac{2\pi p_{0}}{a^{2}}\left(
\ln\frac{\xi^{*}}{a}+\gamma\right)e^{-a/\xi^{*}},
\label{Eq:A1}
\end{eqnarray}
where $p_{0}$ is the size of momentum at the nearest neighbor site of the origin which is assumed to 
be the impurity ($\mu^{+}$) site, and $\gamma\simeq 0.577\cdots$ is the Euler constant.
}

\end{document}